# Report on Community Cadence Observing to Maximize the Scientific Output of the Keck Planet Finder

**Co-Chairs:**
Erik Petigura (UCLA) and Andrew Howard (Caltech)

**Committee Members:**
Jacob Bean (U. Chicago), Charles Beichman (NExScI),
Debra Fischer (Yale), BJ Fulton (NExScI), Howard Isaacson (UC Berkeley),
John O'Meara (WMKO), Carolyn Jordan (WMKO), Daniel Huber (UH),
Paul Robertson (UC Irvine), Arpita Roy (STScI), Johanna Teske (Carnegie),
and Josh Walawender (WMKO)

# Executive Summary

The arrival of the Keck Planet Finder (KPF) in 2022 represents a major advance in the precision radial velocity (PRV) capabilities of the W. M. Keck Observatory. KPF’s cadence of observations will play a central role in its scientific return and by extension continued scientific excellence of Keck Observatory. In preparation for KPF science, our committee of PRV experts and WMKO staff studied the current implementation of cadence observing at Keck and other PRV facilities. We find that many of KPF’s major science cases are not feasible through Keck’s standard allocations of full or half nights to individual PIs. Pooling time among several PIs as is currently done by the California Planet Search (CPS) collaboration with HIRES results in lower quality science results than is possible when KPF is available at higher observational cadence. The CPS strategy was effective for an instrument with poorer Doppler precision than KPF and for a scientific era where the Doppler signals of interest dominate over the noise sources. This strategy also creates barriers to entry, particularly for researchers wishing to lead small proposals.

This report makes recommendations for optimizing PRV cadence at Keck subject to the following constraints: preservation of clear boundaries between cadence observations and classically scheduled time; and ensuring fairness and scientific independence of different Keck TACs and different KPF PIs. We recommend establishing a new category of Keck time allocation, “KPF Community Cadence” (KPF-CC). KPF users would opt into KPF-CC at the time of proposal submission. In many ways, KPF-CC will formalize observing strategies provided by CPS, but with higher observational cadence appropriate for KPF science and with universal access to the program for all Keck users. We recommend that KPF-CC time be scheduled classically into blocks as small as a quarter night subject to considerations of bright/dark time, variations in proposal pressure with the seasons, and the needs of non-KPF observing programs. Within KPF-CC time, the Keck Observing Assistants would execute observations generated by a dynamic scheduler. We recommend that Keck staff and a board of PRV experts design and maintain the scheduling software. Our recommendations for KPF-CC build on decades of experience of CPS observations using HIRES at WMKO as well as practices developed more recently at facilities with instruments having Doppler precisions that are comparable to KPF’s.

We view KPF-CC in the spirit of other organizing principles of the Keck schedule that our community has come to embrace, including scheduling dark time observations where scientifically justified, date-specific observations for timed celestial events, and allowing for interrupts for target-of-opportunity observations.

# Introduction

The Keck Planet Finder (KPF) is a high-resolution (R = 95,000) echelle spectrometer with a bandpass covering 445-870 nm that will see first light on Keck I in summer 2022. KPF is designed for precision radial velocity (PRV) measurements of 30 cm/s. Key technologies that contribute to KPF's PRV capabilities are

- an all-Zerodur spectrometer design for extreme thermomechanical stability;
- an illumination stabilization system with octagonal cross-section fibers, optical scrambling, active fiber agitation, and image slicing;
- an accurate and precise wavelength calibration system through the combination of a laser frequency comb (LFC) and a stabilized etalon;
- and high optical efficiency (~7.5% at peak).

A PRV system's scientific productivity depends on:

- Instrumental velocity stability
- Efficiency
- Time awarded
- Cadence – the distribution of awarded time

In terms of instrumental stability and RV efficiency, KPF represents significant advances relative to its predecessor instrument HIRES; we compare the two instruments in Table 1. Note that while KPFs advantage in *raw optical efficiency* is modest (5.7%→7.5% in nominal conditions), its *Doppler efficiency* outperforms HIRES by a factor of ten by eliminating the need for spectral forward modeling with the iodine cell. Future TACs will decide the time available for KPF science, but as a point of reference, PRV science with HIRES received a total of 45 and 48 nights in the 2020B and 2021A semesters. The cadence of KPF observations has yet to be determined and is the subject of this report.

| | **HIRES** | **KPF** |
|---|---|---|
| **Velocity stability** (systematic noise floor) | 2 m/s | 0.3 m/s |
| **Efficiency** | Photon: 5.7% at peak<br>RV: ~20 min to achieve 2 m/s for V = 12 mag | Photon: 7.5% at peak<br>RV: ~2 min to achieve 2 m/s for V = 12 mag |
| **Time awarded** | Varies by semester; pooled by CPS.<br>2020B: 48 nights<br>2021A: 45 nights. | TBD by future TACs |
| **Cadence** | Combination of 25, 50, 75 and 100% nights, executed by CPS.<br>2020B: 70 full/fractional nights<br>2021A: 64 full/fractional nights | TBD by Keck SSC |

*Table 1. Comparison of HIRES and KPF capabilities.*

KPF differs from other Keck instruments in having a single optical input, a lack of moving parts in the spectrometer, and a lack of other adjustable features common to general-purpose instruments. KPF will be calibrated automatically every morning and afternoon to track the evolution of instrument drift whether it is scheduled for observations that night or not. These features of simplicity and automated operations facilitate the cadence observing style that we propose here.

Funding for KPF came from a variety of sources, but the biggest single award was a $6M NSF MSIP grant. In their proposal, the instrument team identified four key science projects with KPF. (Of course, we anticipate many other KPF projects with varying sizes and science topics.) The four key cases were:

1. Discovery/characterization of super-Earths orbiting nearby Sun-like stars. In parallel, chart a path to 10 cm/s through a deep study of stellar Doppler variability.
2. Discovery/characterization of Earth-mass planets orbiting M-dwarfs. Optimal targets for direct imaging and spectroscopy by GSMTs.
3. Measure the bulk densities and compositions of hundreds of TESS-discovered planets.
4. Study the densities and compositions of the Earth-size planets from Kepler.

All of these science cases require observations on many nights to sample orbital phase. Moreover, for most stars, stellar phenomena such as acoustic oscillations, convection, rotating surface features, and magnetic activity are expected to produce RV signals that exceed the instrumental stability of KPF. High cadence observations are required to properly model such noise sources.

To study cadence opportunities for KPF, two of us (Erik Petigura and Andrew Howard) were invited by the Keck SSC to form a committee of PRV experts and WKMO staff. This is a report of the committee's findings and recommendations, organized as follows: We describe the committee makeup, expertise, and activities (section: "KPF Cadence Committee"). We expand upon several scientific motivations for cadence (section: "Scientific Motivation for Cadence"). We survey other PRV facilities and summarize various approaches to cadence (section: "Current Cadence Solutions"). We present a set of policy recommendations for WMKO (section: "Policy Recommendations"). We show that these policy recommendations are compatible with a realistic Keck semester using a mock semester schedule (section: "Proof of Concept"). We sketch out an implementation plan, estimate costs, and recommend future work (section: "Next Steps"), and provide a brief conclusion (section: "Conclusion").

# KPF Cadence Committee

## Committee Membership and Expertise

In April 2021, we formed a committee to study cadence observing at Keck and other PRV facilities and to develop recommendations for optimizing KPF's cadence capabilities, while interfacing well with other instruments, the observatory's existing structure, and the time allocation process. We worked to include expert users of Keck/HIRES and other PRV facilities, researchers with experience with both classical and queue observations, and WMKO staff who are key to the operations and scheduling of KPF and Keck. We list the committee members and their relevant expertise below:

| **Person** | **Role/Expertise** |
|---|---|
| Erik Petigura, UCLA | Co-chair |
| Andrew Howard, Caltech | Co-chair, KPF PI and CPS leader |
| Johanna Teske, Carnegie | Leads Magellan/PFS RV scheduling |
| Chas Beichman, NExScI | Provides perspectives from NASA partnership |
| Josh Walawender, WMKO | Keck Staff Astronomer, KPF instrument lead |
| John O'Meara, WMKO | Keck Chief Scientist |
| Jacob Bean, U. Chicago | PI of Gemini/MAROON-X, experience with Gemini queue |
| Paul Robertson, UC Irvine | Experience with HET/HPF and WIYN/NEID, provides perspectives from UC |
| Carolyn Jordan, WMKO | Keck scheduling lead and Observing Assistant |
| Dan Huber, UH | Current Keck/HIRES user, provides perspectives from UH |
| Debra Fischer, Yale | PI of DCT/EXPRES, provides perspectives from a small partner institution |
| Howard Isaacson, UC Berkeley | Operations lead for CPS, 400 nights observing experience with HIRES |
| BJ Fulton, NExScI | Developed robotic scheduler for APF/Levy |
| Arpita Roy, STScI | KPF Project Scientist |

## Timeline of Committee Work

- Mar. 2021 – At the Keck SSC meeting, Petigura and Howard gave a presentation on the status of KPF and the need for cadence observations to realize the science goals.
- Apr. 2021 – Committee formed
- Apr-Aug. 2021 – Committee met twice per month and solicited presentations from representatives from WMKO and facilities to document the range of approaches to observing cadence and scheduling.

- Andrew Howard on cadence with Keck/HIRES
    - Heidi Schweiker & Mark Everett on cadence at WIYN/NEID
    - Carolyn Jordan on how the Keck semester schedule is made
    - Debra Fischer on cadence at SMARTS/CHIRON and DCT/EXPRESS
    - Johanna Teske on cadence at Magellan/PFS
    - Xavier Dumusque on cadence at ESO/HARPS and VLT/ESPRESSO
    - Jacob Bean on cadence at Gemini/MAROON-X
- Sep. 2021 – Committee synthesized findings and proposed recommendations
- Sep. 2021 – A subgroup (Jordan, Petigura, and Howard) created a mock 2021A schedule implementing the proposed recommendations.
- Sep-Oct. 2021 – A subgroup (Fulton, Petigura, Howard, Robertson, Isaacson) created mock datasets according to different observing cadences.
- Oct-Nov. 2021 – Full committee provided feedback on the report.
- Nov. 2021 – Petigura and Howard present at the Keck SSC meeting

## Scientific Motivation for Cadence

A number of science cases for KPF are optimized through small amounts of time distributed across many nights. Most science cases are **not possible** given classical awards of full or half-nights. The common thread through these science cases is that the effects of stellar activity limit Doppler sensitivity. Time-variable surface features produce shifts in the stellar spectra that register as Doppler shifts that have nothing to do with the star's motion around the star-planet barycenter. The amplitude of activity RVs ranges from tens of cm/s for the very quietest stars to hundreds of m/s for young and active stars. The timescale of this stellar variability ranges from minutes (acoustic modes) to hours (granulation), to days (rotation), to years (magnetic cycles). Rotationally-modulated activity signals are especially challenging to deal with; these quasi-periodic noise sources cannot be binned down like random noise and must be sampled in the time domain and modeled to effectively remove the signals. The majority of the stars will have activity amplitudes exceeding KPF's instrumental precision of 30 cm/s. Thus the smallest detectable Doppler signals will be limited by stellar activity unless it can be modeled out. Activity modeling requires high cadence and we present several scenarios below.

Several strategies have been developed for mitigating stellar activity. Most depend on modeling the time series of RVs as the sum of Keplerian contributions (from orbiting planets), quasi-periodic contributions (from stellar activity, especially rotationally-modulated activity), and random noise (from photon noise and instrumental effects). The quasi-periodic part of the model is often a Gaussian process (e.g., Rajpal et al. 2015) or a physically-inspired model (Aigrain et al. 2012). In both cases, *the model depends on sampling the activity signal many times over its stellar rotation timescale.* These models are trained on activity indicators that can be derived from the echelle spectra themselves (e.g., $\log R'_{HK}$ or statistics of the cross-correlation function) or contemporaneous photometry. The key is to separate the Fourier components of the planet model from Fourier components of the stellar activity model. Importantly, methods that employ a linear decorrelation of RVs with an activity metric are significantly less effective at mitigating activity than time series modeling with high cadence observations (Siegel et al. submitted).

Thus, high cadence RV observations are needed to remove the clouding effects of stellar activity; sporadic RV measurements are intrinsically limited in their ability to suppress stellar activity and probe the smallest planets. Figure 1 shows examples from the literature of high-cadence measurements from KPF-like instruments being used to suppress stellar activity.

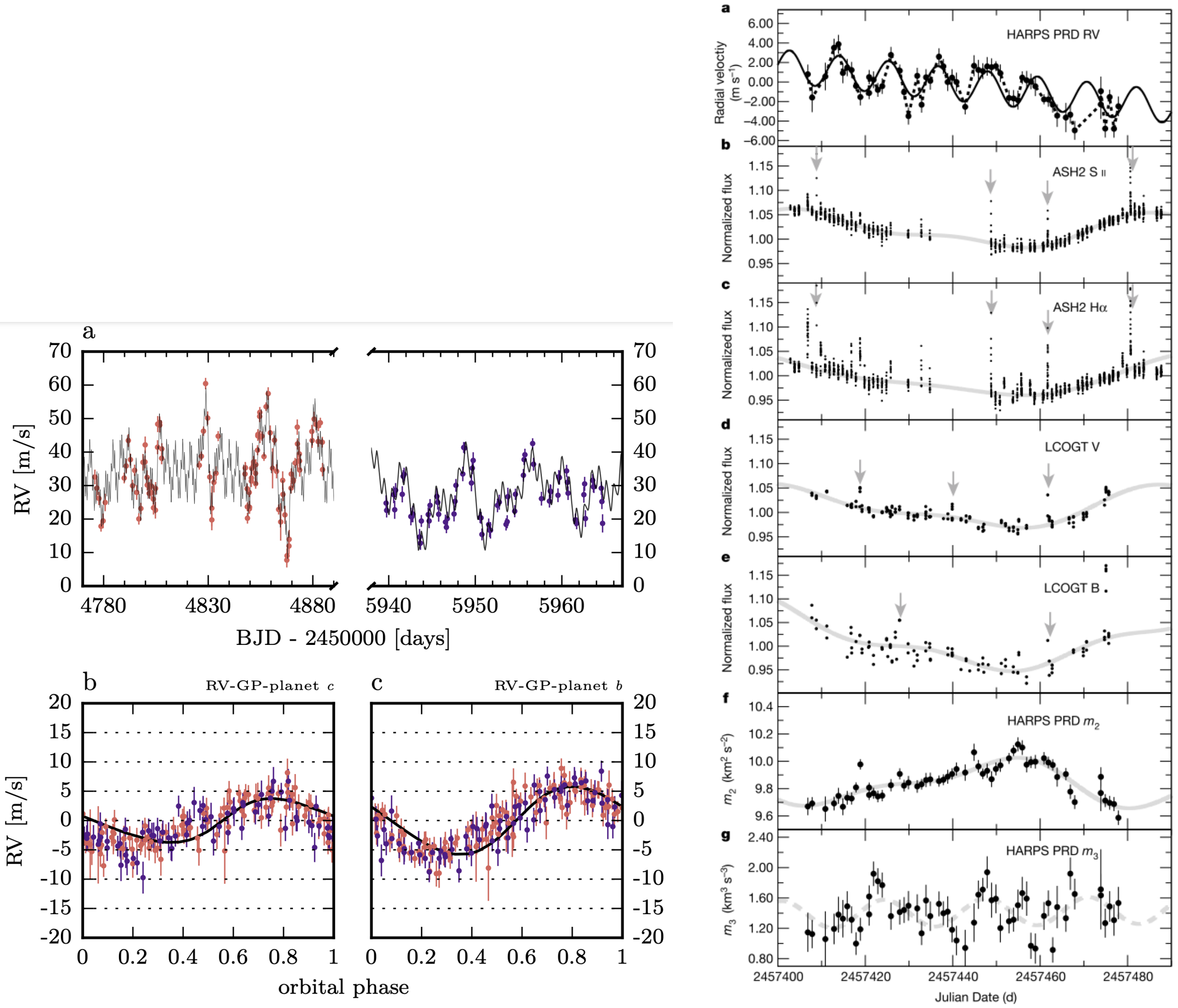

*Figure 1. Examples of high cadence RVs used to mitigate stellar activity from the literature. Left: Faria et al. (2016) measured the masses of the transiting planets Corot-7b and Corot-7c using a Gaussian process model for stellar activity (without the help of photometry). The relevant timescales here are 0.5-day and 3.7-day orbital periods for the planets and a 23-day rotational period of the star. Right: the discovery of Prox Cen b (a planet with a mass of 1.3 Earth-masses in the habitable zone of the nearest late M dwarf) in HARPS RVs by Anglada-Escudé et al. (2016) depended on modeling out a complex activity signature tracked by spectroscopic indicators and photometry.*

The importance of achieving observational cadence to model stellar activity is widely appreciated in the Doppler planet-hunting community. For example, the final report of the Extreme Precision Radial Velocity Working Group (Crass et al. 2021) concludes that an “inability to obtain sufficient observing time/cadence” is a significant risk factor in Doppler searches for small planets that should be mitigated.

## Examples of Activity-Dominated KPF Science Cases

### 1. Discovery/characterization of terrestrial planets in the habitable zones of M-dwarfs

For M-dwarfs the liquid-water "habitable zone" (HZ) corresponds to orbital periods of ~5-40 days depending on their mass. To detect (blind discovery) or characterize (e.g., measure mass) a terrestrial-mass exoplanet with these periods, one must detect reflex motion with semi-amplitudes $K \lesssim 2$ m/s. For middle-aged or old M dwarfs, the dominant source of stellar RV noise is spots modulated by the rotation period which is typically ~1 month or more. In many cases, the orbital period is shorter than the stellar rotation period. Thus, the most reliable way to recover the planet is to observe at high cadence, resolving the high-frequency planet signal atop a slow drift created by stellar activity. This technique was implemented effectively in the discovery of an Earth-mass planet in the HZ of Proxima Centauri (Anglada-Escude et al. 2016). Nightly observations on HARPS were critical to recover the $K = 1.4$ m/s reflex motion in the presence of rotational modulation that was ~2 m/s (see Figure 1).

### 2. Mass measurements of super-Earths in the habitable zones of Sun-like stars.

The Kepler Space Telescope detected a number of terrestrial-size planets in the HZs of GK stars (e.g., Kepler-186f, Quintana et al. 2014). For such stars, the inner edge of the HZ corresponds to $P \sim$ 100-400 days. While true Earth-analogs with $K \approx 10$ cm/s lie below KPF’s target first light instrumental noise floor of 30 cm/s, it is possible that, in coming years, we will understand instrumental systematics at the level required to detect these small signals. In the near term, KPF’s large aperture and instrumental stability will allow Keck astronomers to measure the masses of slightly larger and more massive “super-Earths” if activity can be mitigated (an $R \approx 1.4\ R_E$ planet is 4 $M_E$ with $K \gtrsim 40$ cm/s). Such systems present two major observational challenges: (1) the RV amplitude of the planet will be significantly smaller than the activity amplitude, and (2) the planet's orbital period will be much longer than the stellar rotation period, typically 20-40 days. Thus, one must resolve and model multiple rotation cycles of a quiet, Sun-like star in order to extract the smaller signal of an exoplanet. The challenge to these measurements will be similar to that faced by Dumusque et al. (2012) in their effort to detect low-mass exoplanets orbiting alpha Centauri B. This study demonstrated why high-cadence observations are essential for the task. First, it is crucial to resolve and model the stellar rotation signal in the time series rather than decorrelating the RVs with activity tracers (which leaves residual activity noise) or whitening the time series in Fourier space (which creates aliasing issues; Rajpaul et al. 2016). Furthermore, since the planet's orbit is comparable to a single observing season, it is best to obtain enough RVs to confidently detect both the planet and any activity signals within each individual season. Otherwise, season-to-season activity level changes (as starspots appear, evolve, and decay) and instrumental baseline drifts make accurately modeling the time series much more challenging (Hatzes et al. 2013).

### 3. Mass measurements of planets around active young stars

Measuring the masses of Neptune- and Saturn-mass exoplanets transiting young, active stars is important for constraining models of planet formation and evolution (e.g., Stefansson et al. 2020). However, the starspot modulation of even moderately young stars (age ~ 0.1–1 Gyr) can induce RV signals with amplitudes of 10–100 m/s (Hillenbrand et al. 2015), while the planetary reflex motion is typically 5–30 m/s. There are two promising techniques for recovering the planets despite such intense RV noise in M dwarfs of this type, both of which require high-cadence sampling. First, for very rapidly rotating M dwarfs, starspots may persist for a full observing season before decaying (Robertson et al. 2020). Thus, with high-cadence observations over a single season, one may "freeze" the starspot signal in place, greatly simplifying the modeling required to detect the planet in the residuals of the rotation signal. For hotter stars, it has been shown that photometric variability from space-based photometers such as Kepler and TESS is highly predictive of RV variability caused by starspot modulation. This technique is known as the "F F-prime" or FF′ method (Aigrain et al. 2012). Thus, if we conduct a high-cadence RV campaign concurrently with TESS observations of a star, we may use FF′ predictions to simply subtract the starspot signal from our RVs, rather than relying on multi-dimensional correlated noise models. Continued operation of TESS and, later, PLATO, make this possibility feasible for years to come (Burt et al. 2018).

## Simulated Planetary Retrievals from Activity-Dominated RV Timeseries

Completing the science projects described above requires sampling and modeling both the planetary and activity RVs. To explore the role that cadence plays in achieving these science goals, we considered a nominal Keck award of 10 nights spread out over up to 4 semesters according to the three "Cadence Modes." For this nominal survey, the PI observes 10 targets with $V$ = 10 a total of $N$ = 60 times. Each exposure is 10 min and achieves photon-limited RV errors of 30 cm/s. *Note that we are considering the same total time = (10 targets) x (60 exposures/target) x (10 min/exposure) = 100 hours, but divided different ways.* To simplify matters, we assume all targets are at the same RA/Dec. As shown in Figure 2, the three cadence modes are:

- **Mode A. "Classical"** — PI is awarded time as 20 half-nights. To generate sample nights, we randomly drew 20 nights over four semesters subject to the constraint that the targets are observable. The observing baseline is 550 days. To achieve 60 observations, this PI must execute three exposures per night.

- **Mode B. "CPS"** — PI pools their time with other KPF PIs in an *ad hoc* queue analogous to the California Planet Search's (CPS) queue, which operates 60 (partial/full) nights per semester or 33% of all nights. We generated sample nights by drawing from the CPS nights scheduled during 2021A until we reached 60. The observing baseline is ~180 days. Each night has a single observation.

- **Mode C. "Community"** — PI pools their time with other KPF PIs in a queue that is on-sky 135 nights per semester or 75% of all nights. We generated sample nights by drawing from the mock 2021A schedule described later in this report. Each night has a single observation.

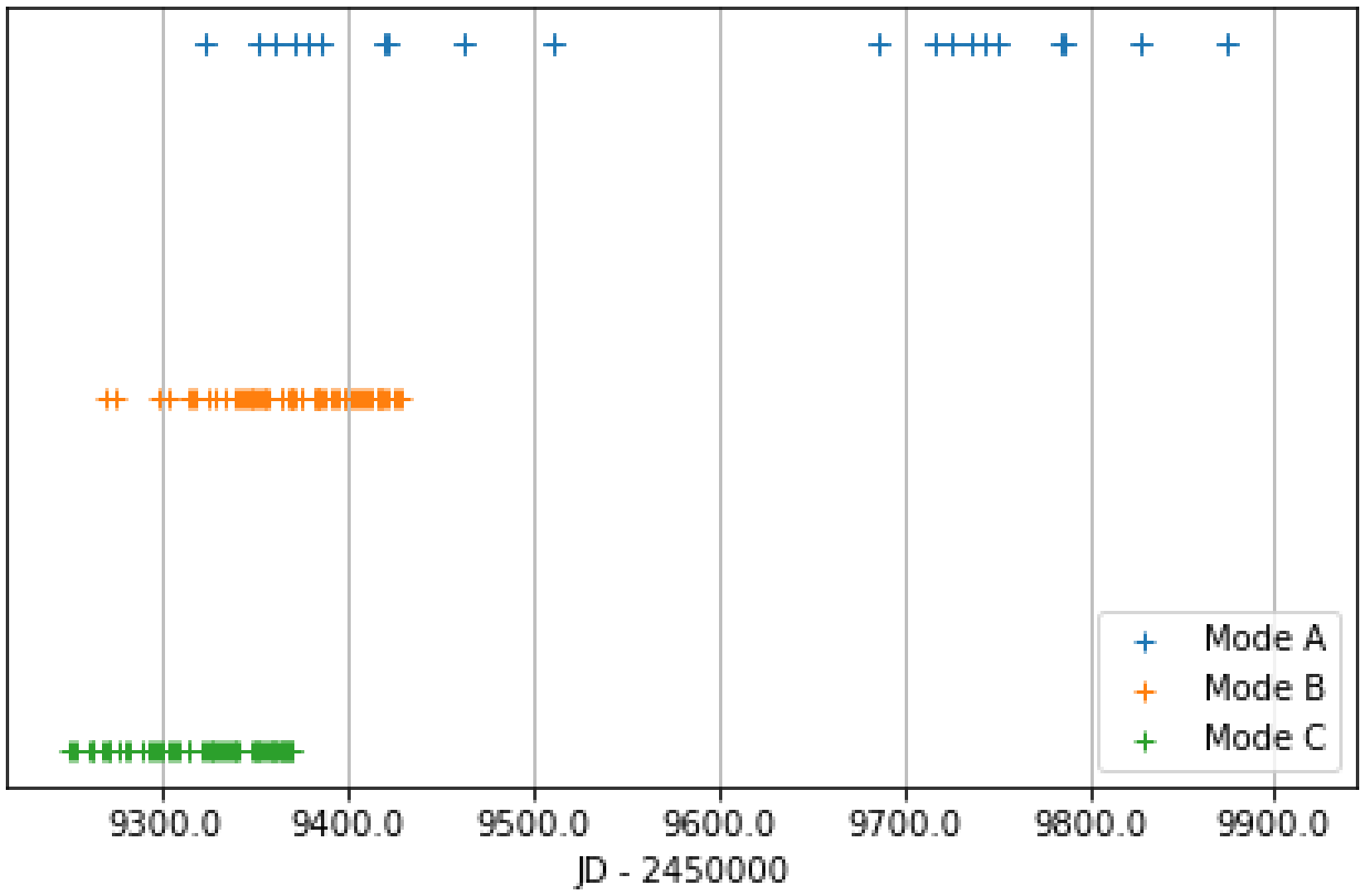

*Figure 2. Distribution of time according to different cadence modes.*

We then generated a synthetic KPF time series with simulated stellar noise, photon-limited noise, KPF instrumental noise, and planetary reflex motion.

1. **Stellar activity.** Gaussian processes (GPs) are the standard statistical description of stellar variability in the exoplanet community (see, e.g., Haywood et al. 2014). GPs can describe a wide range of correlated noise by specifying the correlation between points of different time offsets through a covariance matrix. The parameters of the convolution kernel that specify the covariance matrix are called hyperparameters.
    a. We modeled granulation noise with a quasi-periodic GP with an amplitude of 0.5 m/s and a periodicity of 1 hr.
    b. We modeled RV noise from the rotation of spots and plage as a quasi-periodic GP with an amplitude of 3.0 m/s and periodicity of 20 days

2. **Photon.** We added white Gaussian noise with σ = 0.30 m/s.

3. **Instrument.** We added white Gaussian noise with σ = 0.30 m/s.

4. **Planet.** We injected a planet on a 4.2-day, circular orbit with a semi-amplitude $K$ = 1.0 m/s. This corresponds to a 2.2 Earth-mass planet orbiting a 1.0 solar-mass star. Note that the planet's orbital period is shorter than the rotation period and the planet has a smaller RV amplitude than the stellar noise.

Figure 3 shows the rotation, granulation, and combined stellar activity. We also show the power spectral density, which approximates that of inactive sun-like stars (Dumusque et al. 2012).

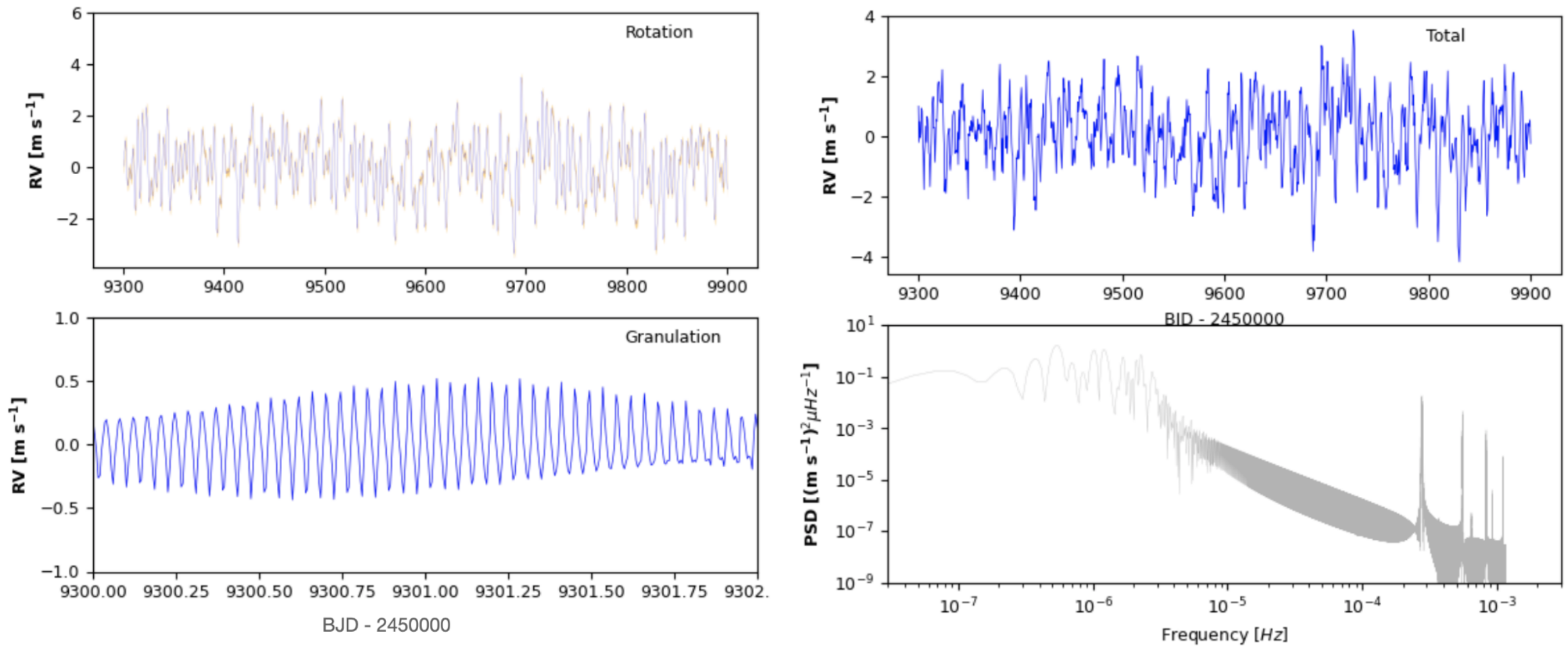

*Figure 3. Upper left: simulated stellar variability from rotation. Lower left: same, but for granulation. Upper right: same but rotation and granulation. Lower right: Power spectral density of activity RVs RVs.*

We simulated the retrieval of this planet's RV signal by simultaneously modeling the time series with a Keplerian and GP. For real systems, we do not know the exact covariance structure of the stellar noise but observationally-motivated priors may be placed on the hyperparameters that describe the GP covariance structure. We modeled the synthetic KPF observations with a *RadVel* standard RV modeling toolkit (Fulton et al. 2018) while imposing realistic priors on the hyperparameters.

Figure 4 compares the planet retrievals under different cadence modes. Figure 5 shows how the significance of the derived planet mass grows as a function of the number of observations and as calendar nights. We compare and contrast the outcomes below:

- **Mode A** uses three, back-to-back exposures, a.k.a. "triple-shot" exposures. However, noise on intra-night timescales is not white, so triple-shot exposures do not result in 1/sqrt(N) reduction of errors. Also, there are typically 1 or 2 sets of triple-shots per rotation period. The cadence is insufficient to build a model of the stellar rotation-modulated variability. **Outcome: PI achieves an apparent 2σ mass measurement in 550 days of observing.** In addition to being imprecise, this measurement is inaccurate. The algorithm recovered a mass of 4.2 ± 1.3 $M_{Earth}$ while the simulation injected a mass of 2.0 $M_{Earth}$. Similarly, the recovered eccentricity was 0.24 while the injected signal was circular. **With sparse sampling, it is not possible to simultaneously model both the planetary and stellar signals.**

- **Mode B** has two advantages relative to Mode A. The larger number of available nights allows the PI to achieve 60 observations in a single observing season with one exposure per night. Each observation has more power as it samples different granulation cycles. In addition, there are many observations per rotation period which allow for better constraints of the GP that models the stellar activity. **Outcome: PI achieves a 8σ mass** **measurement in 180 days.**

- **Mode C** has even higher sampling density than Mode B and results in a more precise model stellar activity. **Outcome: PI achieves a 10σ mass measurement in 120 days.** The orbital eccentricity is also constrained with 3× the precision in Mode C compared to Mode B.

## KEY FINDING

Our simulations and many previous studies in the literature show that high cadence observations are critical to PRV science when planetary signals are comparable to stellar activity. For our nominal planet, Mode A does not result in a scientifically compelling mass measurement. The mass measurement with Mode B would allow for powerful constraints on planetary composition, but science return per Keck night is highest for Mode C.

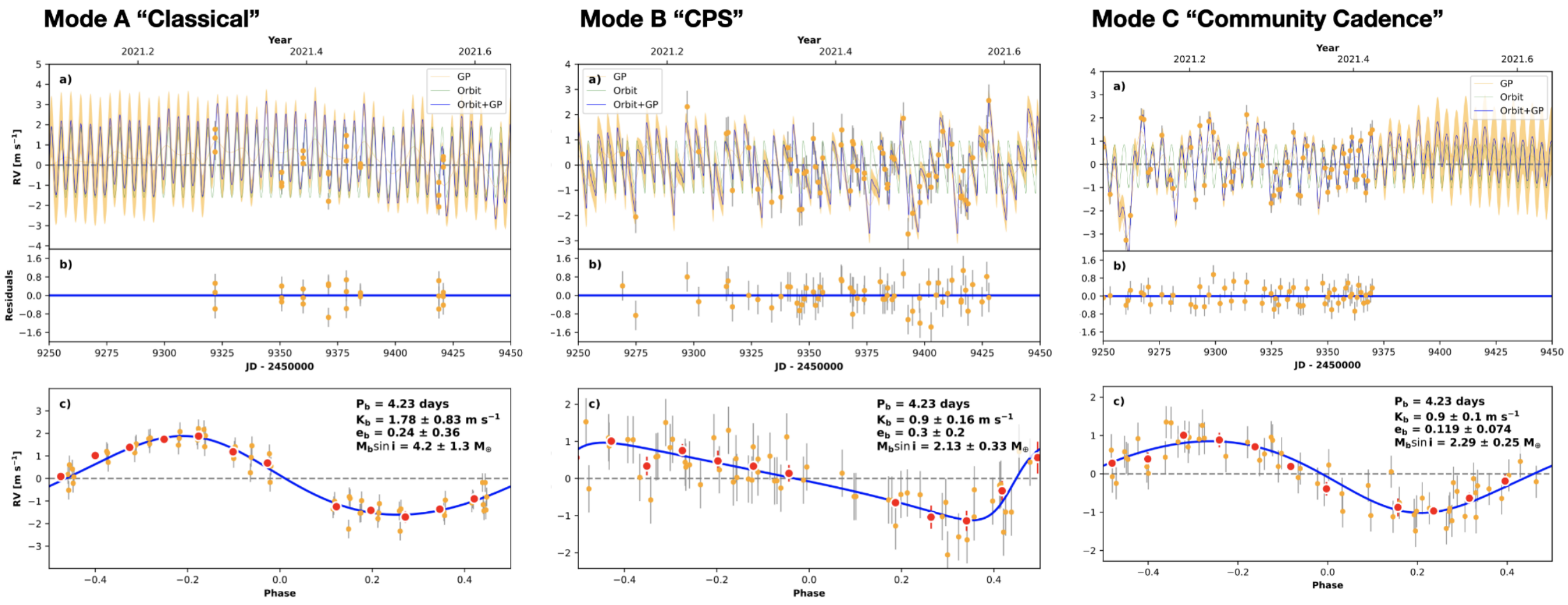

*Figure 4. Left top: points show synthetic KPF RVs observed according to Cadence Mode A. The full observing baseline is 550 days, but we show 180 days for clarity. The blue line shows the best-fit model (technically, the maximum a posteriori “MAP” model). It is the sum of planetary and stellar components, shown as green and yellow lines. The 1σ uncertainty on the combined model is shown as a yellow band. The width of the band is ~1 m/s and limits the power of each isolated triple-shot exposure to constrain the planet model, and we achieve a 2σ mass measurement. Left bottom: the blue line shows the phase-folded MAP orbit model, and the yellow points show phase-folded RVs after subtracting out the MAP activity model. The dispersion of the yellow points about the blue line may seem to suggest that we should have measured the mass to better than 2σ. However, we have only subtracted out one of many credible noise models. To properly account for the uncertainty in the GP activity model, we have marginalized over all credible activity models. Middle: same as left but for Cadence Mode B. The uncertainty in the activity model decreases to ~0.25 m/s in the high cadence region and reverts to ~1.0 m/s in regions of sparse sampling. We achieve an 8σ mass measurement. Right: same as left, but for Cadence Mode C. The uncertainty in the activity model is smaller than even Mode B. We achieve a 10σ mass measurement.*

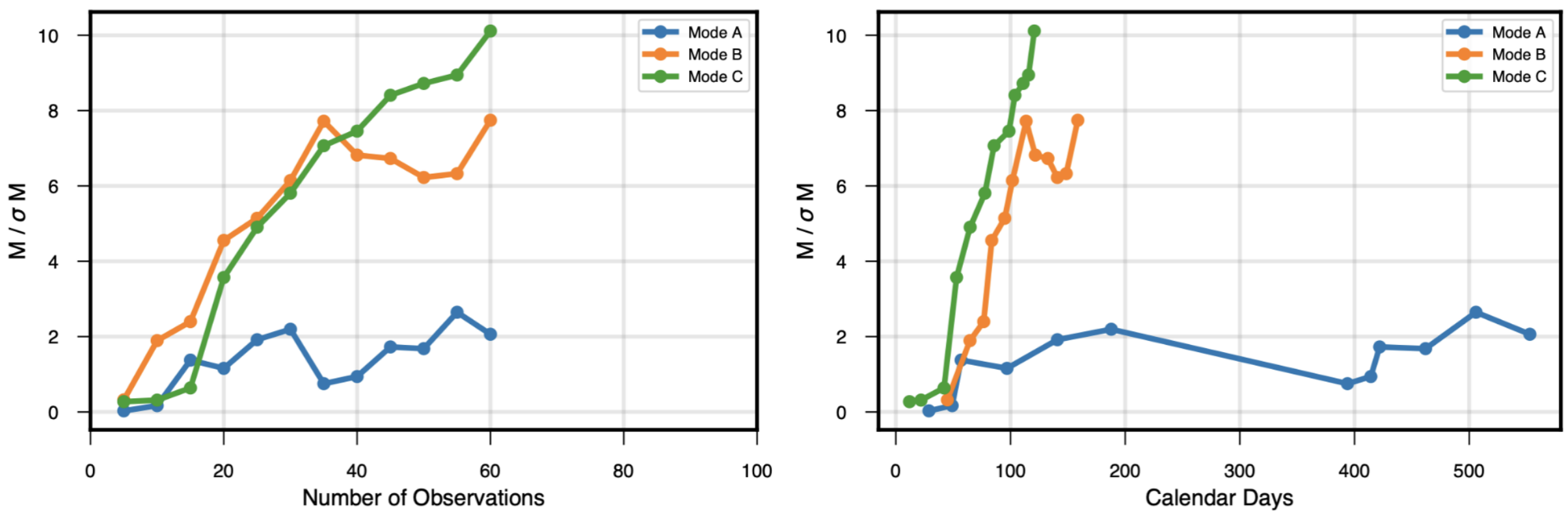

*Figure 5. Left: injected planet mass M divided by the uncertainty on the modeled M as a function of the number of observations. The significance of the recovered M grows most rapidly with mode C. Right: same as left but with calendar nights on the x-axis.*

## Current Solutions to Cadence

A key activity of our committee was to survey the landscape of PRV cadence observing in the exoplanet community. We solicited presentations from a number of PRV groups on how they address their cadence needs. Specifically, we asked each group to address the following topics:

1. **Awarded time**: What is the total number of nights and their calendar distribution?
2. **Cadence implementation**: How are nightly observing scripts generated? How are observing priorities factored in? How are observations taken? Who takes them? How are RVs reduced/distributed? How are observing priorities factored in? How much collaboration is required among queue participants?
3. **Policy**: How are competing science interests/target overlap adjudicated? Where does decision-making authority lie?
4. **Effort (FTEs)**
5. **Pros**
6. **Cons**

The table below compares these different instruments. They are ordered by instrument first light date.

| | |
|---|---|
| Keck/HIRES<br>Andrew Howard | • **Instrument summary:** Slit-fed, non-stabilized spectrometer with iodine cell at 10 m Keck 1 telescope. First light: 1994. Single exposure noise floor of ~2 m/s.<br><br>• **Awarded time:** PRV science at HIRES has historically received 45-50 nights per semester which is distributed over 60-70 full/fractional nights for the |
| | California Planet Search (CPS) collaboration. The calendar distribution is driven by requests on WMKO coversheets. Typically observers request two high-cadence runs per semester and no fewer than one night per lunation. Generally, the awarded time has been anchored by a small number of large, multi-semester projects and a larger number of smaller 1-3 night projects.<br><br>• **Cadence implementation:** A. Howard leads CPS which manages an *ad hoc* queue. Team management, data reduction, curation, quality control, preliminary analysis are led by H. Isaacson. Participating groups contribute observers (currently ~25) who are trained by experienced CPS observers in ~10 nights.<br><br>• **Policy**: CPS members are colleagues and collaborators with a relationship built on trust. The CPS collaboration is open to collaborations across the Keck partnership and typically measures RVs for 10-12 science programs per semester. CPS aims to adjudicate overlapping scientific or target interests collegially. In cases of irreconcilable overlap, science/target priority goes to core CPS members and their students. There are cases where CPS has declined collaboration when the proposed science substantially overlaps with an existing CPS project. Decision-making authority rests with core CPS members and finally with A. Howard. This *de facto* approach to cadence observing with HIRES creates a barrier to entry for some members of the Keck community.<br><br>• **Total effort:** 1 FTE (excluding observers). Mostly H. Isaacson and A. Howard.<br><br>• **Pros**: Facilitates scientific collaborations; lightweight bureaucracy; experiments, pilot projects, and observing favors are encouraged.<br><br>• **Cons**: PIs who wish to perform PRV science with HIRES are forced to collaborate with CPS. Significant CPS effort for community benefit. There are modest benefits to CPS supporting smaller programs, the smaller programs would often not be possible without the support of CPS. |
| HARPS<br>Xavier<br>Dumusque | • **Instrument summary:** fiber-fed, stabilized spectrometer at 3.6 m La Silla telescope. Precision 1 m/s. First light: 2002.<br><br>• **Awarded time**: Individual PIs propose for time through ESO. Usually ~10 PIs have been allocated ~70-110 nights/semester. Time is distributed throughout each semester in ~16 runs of ~2-8 days each.<br><br>• **Cadence implementation**: Cadence is achieved through an ad hoc queue led by X. Dumusque and F. Bouchy. To simplify scheduling only 50% of programs are executed within a given run. X. Dumusque synthesizes observing requests from different PIs in advance of the night and generates a script. Time allocation and program completion is tracked on a Google spreadsheet. The total number of targets is ~20-30 per night. Participating |
| | teams supply observers who are trained in ~2 nights.<br><br>• **Policy**: PIs who want to make use of the queue are required to collaborate on duplicated targets. It is not possible to achieve cadence outside this queue, so direct competition between HARPS users and target duplication is avoided.<br><br>• **Effort:** 4 weeks of time per year for X. Dumusque and F. Bouchy<br><br>• **Pros:** Despite the considerable coordination effort, Dumusque believes this is a highly efficient way of using the telescope and plans to continue this mode into the future.<br><br>• **Cons:** Nightly cadence is not possible. |
| Magellan/PFS<br>Johanna Teske | • **Instrument summary:** Slit-fed, non-stabilized spectrometer at 6.5 m Magellan telescope. Precision: 1 m/s. First light: 2010.<br><br>• **Awarded time:** 30-50 nights/semester, scheduled in 1-2 week runs. ~50% of the time is awarded by Carnegie TAC.<br><br>• **Cadence implementation:** Cadence is achieved through an ad hoc queue managed by J. Teske. PIs submit target information to Teske 1-2 weeks before each run. Each star and its metadata (e.g., finder charts) gets input into a Trello card for organization. Typically there are 15-20 stars per run and 40 stars observed through the semester. The focused effort of individual PIs on a small number of stars sometimes leads to high RA pressure during some parts of the night. For each run, Teske creates a block schedule for both good and bad conditions. Teams that participate in the queue are encouraged to supply observers but most do not. After the run, data reduction and distribution is performed by P. Butler. An overhead is charged for good seeing. Weather and overhead are shared equally in proportion to time contributed.<br><br>• **Policy:** PFS is a PI instrument, so scientists must collaborate with the PFS team. PIs with overlapping science interests are required to collaborate.<br><br>• **Total effort:** 0.5 FTE. Mostly J. Teske<br><br>• **Pros**: Informal. In a given run, almost everyone gets some data.<br><br>• **Cons**: Significant amount of effort to manage the queue. Current personnel cannot support more than 60 nights per semester. Human bias can affect scheduling. |
| VLT/ESPRESSO<br>Xavier Dumusque | • **Instrument summary:** Fiber-fed, stabilized spectrometer at 8 m VLT. Can be fed by any/all UTs, but 4-telescope mode rarely used for exoplanets. On-sky precision of 30 cm/s. First light: 2017. |
| | • **Awarded time**: GTO program is 273 nights / 4 years or 34 nights / semester.<br><br>• **Cadence implementation**: 50% of GTO time is executed in "service mode." ESO staff astronomers conduct these observations. They receive a 20% surcharge but are guaranteed. Scheduling is done by VLT staff. ESPRESSO makes effective use of small (~30 min) blocks of time on different UTs. 50% of GTO time is "visitor mode" (similar to classical). The team uses this for transit observations.<br><br>• **Policy:** GTO time is divided into three working groups each with a chair.<br><br>• **Effort:** Unknown FTE. Done by ESO<br><br>• **Pros**: Service mode has been effective for high-cadence observations such as observing Proxima Centauri twice per night for a semester. Visitor mode works well for transit/Rossiter-McLaughlin observations.<br><br>• **Cons**: Long-term RV monitoring for a large number of stars such as the nearby planet search program has proved challenging given the complexity of the ESO queue system. Dumusque said this works much better at La Silla / HARPS where collaboration controls ~50% of telescope time. |
| WIYN/NEID<br>Heidi Schweiker & Mark Everett | • **Instrument summary:** Fiber-fed, stabilized spectrometer at 3.5 m WIYN telescope. Design goal of 30 cm/s precision; demonstrated sub-40 cm/s on solar spectra. First light: 2021.<br><br>• **Telescope time:** 70 nights per semester, or about 50% of science nights at WIYN. NOIRLab sets the calendar distribution but works to ensure that there are no more than two consecutive non-NEID nights. Successful proposals are awarded time at one or more priority levels. Six partner institutions run independent TACs. NN-Explore is the largest with ~50% and five other university TACs allocate fractions of 22% or less.<br><br>• **Cadence implementation:** PIs submit target requests through a NEID portal where they can attach a priority to specify cadence parameters and other criteria for observations like exposure times, maximum cloud extinction, minimum moon separation, etc. WIYN Queue Coordinator runs the scheduling program that returns a schedule for a full or partial night. The scheduling program is complex and is descended from the HET queue. In brief, it schedules long, high-priority, and time-restrictive observations in a first pass. Then, a second pass fills in shorter, lower priority observations into remaining time. Each potential observation is given a ranking based on factors like a program's time balance, slew-acquisition overheads, and whether the target is late in its observing season.<br><br>• **Policy**: TACs are independent so competing science programs are allowed, as are duplicated targets. Once the data is taken, the metadata are public so |
| | competing teams know about the existence of the data, but there is a proprietary period.<br><br>• **Effort:** NEID Instrument Scientist (1 FTE), 1 Queue Coordinator (0.5 FTE), 3 Queue observers (2.5 FTEs), Programers (1.5-2 FTE cumulative).<br><br>• **Pro**: Independent PIs do not need to communicate with one another.<br><br>• **Cons**: The learning curve for many users and TACs is steep. Queued observation requires extensive software development and hand-tweaking is usually required after the scheduler is run. There is some concern that PIs could game the scheduling algorithm to the point that it breaks. |
| MAROON-X<br>(J. Bean) | • **Instrument summary:** Fiber-fed, stabilized spectrometer at 8 m Gemini-N telescope. Precision of 30 cm/s on late M-dwarfs as faint as V = 17. First light: 2019.<br><br>• **Awarded time**: Determined by proposal pressure to NOIRLab. In 2021B, 300 hours was approved at 4:1 oversubscription. M-X was scheduled during 55 nights over three runs of about ~2 weeks each throughout the semester. When the instrument is on the telescope, it conducts ~50% of the observations and generally receives priority over other instruments.<br><br>• **Cadence implementation**: Each afternoon, Gemini staff create several night plans based on forecasted weather, image quality, water vapor, etc. This process is opaque to the M-X team, but it factors in Band ranking (1-4, lower is better), observing conditions (image quality and cloud cover matter for RV, but also sky brightness and water vapor), time-critical nature, and completion status. The schedule is determined to the minute and there is no exposure meter-based cutoff. M-X team takes data although not much input is required and they stand by when the schedule flips to another instrument. The M-X team reduces data and delivers it to the community.<br><br>• **Policy:** NOIRLab awards time and the International Time Allocation Committee adjudicates duplicate targets.<br><br>• **Effort:** At least 1 FTE from Gemini Staff<br><br>• **Pros**: Multiple chances at time-critical observations like transit/Rossiter-McLaughlin, which has been key for early demonstrations of instrument capabilities. High completion rate for Band 1 programs.<br><br>• **Cons**: Not a good track record of completing community programs. Often partially completed datasets are not that useful. |

In summary, the six facilities listed above all implement a queue of some sort. These may be grouped into the following broad categories:

- ***Ad hoc queue.*** *Keck/HIRES, LaSilla/HARPS, Magellan/PFS, VLT/ESPRESSO [visitor mode]* – There is a *de facto* lead group that pools time from different participants. This group ensures cooperation/collaboration on overlapping targets or science interests or may decline to support certain RV science programs. Lead group plans observations, reduces data, and distributes data.

- ***Observatory run queue*** *WIYN/NEID, Gemini-N/MAROON-X, VLT/ESPRESSO [service mode]*) – TAC awards time and the observatory schedules observations. In some cases (e.g. MAROON-X), a single TAC awards all or most of the time and adjudicates overlapping science/targets adjudicated. In others (e.g., NEID) different partner institutions are independent and overlap is possible.

It is worth emphasizing the limitations of the current CPS style of achieving cadence that is used with HIRES. As demonstrated in the above section on simulated data, CPS-style cadence will achieve significantly poorer scientific results for cases where stellar activity signals are comparable in amplitude to the Doppler signals. Most of the frontiers of exoplanet research are in this domain. The poor measurement performance includes significantly lower fractional precision on measured planet mass and orbital eccentricity and longer times needed to complete an observational program. Programmatically, the CPS-style system also has major drawbacks as a *de facto* service for a public observatory. Because the system is managed and executed by a scientific collaboration, obtaining RV measurements with reasonable cadence requires collaborating with that team. While the CPS team tries to accommodate all reasonable requests, some projects are turned down (or perhaps not even proposed) for collaboration with CPS because they overlap with the scientific priorities of core CPS members or their students. While these policies are understandable for a research group expending their own resources to observe and process the data, we assert that all potential KPF users should have access to high-quality observational cadence and data products. (The Data Reduction Pipeline provided by the KPF team will make high-quality data products available to all KPF users and the community at large through KOA.)

# Policy Recommendations

## Guiding principles

Informed by our sampling of cadence solutions above, we considered various solutions for Keck/KPF. To further guide our thinking, we adopted the following guiding principles:

1. Maximize science return from KPF
2. Preserve independence of different partner TACs
3. Preserve scientific independence of different KPF PIs
4. Minimal interference with classically scheduled time
5. Make efficient use of bright time
6. Prioritize simple solutions

7. Specify policy for how
    a. observations will be conducted
    b. data are distributed
    c. data rights and target lists are respected
    d. fair allocation of time is ensured.
    e. target duplications are addressed
    f. issues with program completion are addressed

## Recommendations

1. **Establish an opt-in queue for some KPF observations.** PIs would opt-in on Keck coversheets.

2. **Long, time-critical observations would remain classically scheduled.** Sequences lasting more than 1 hour (e.g., Rossiter-McLaughlin or transit spectroscopy) are a significant fraction of the desired ¼ night KPF blocks and thus lock up the queue and override many shorter observations. Keeping these observations off of the queue, at least initially, will simplify the scheduling software.

3. **Observing assistants conduct queue observations.** OAs do not have an interest in observing one program over another, which promotes intra-program fairness. KPF observations are expected to be very simple because the instrument has a fixed optical format and observing parameters will be presented to the OA and loaded into the KPF observing software through a DSI interface. The OA will also not need to choose between targets (except perhaps in adverse conditions). As described in the Next Steps section below, the burden of executing KPF Community Cadence observations will not overly burden the OAs.

4. **Cadence observations would specify the following properties:**
    a. Coordinates
    b. Range of acceptable airmass
    c. Minimum acceptable moon separation (deg)
    d. Maximum acceptable sky brightness (mag/arcsec^2). Most relevant for faint targets where sky subtraction is a significant term in RV error budget.
    e. Weather quality band. Specified based on throughput relative to a "nominal" (e.g. median value).
        i. Band 1: <1.5x nominal. (e.g., nominal $t_{exp}$ > 20 min.)
        ii. Band 2: 1.5-3x nominal. (e.g., nominal $t_{exp}$ < 20 min.)
        iii. Band 3: >3x nominal. (e.g., nominal $t_{exp}$ < 5 min, very bright targets)
    f. Real-time determination of sky brightness and weather quality band can and should be based on quantitative metrics. The Keck UNO all-sky camera and CHFT MASS/DIMM instrument can provide these metrics.
    g. Exposure settings. Specify seconds or SNR setpoint.
    h. Sampling strategy

i. Requesting observations within defined range in orbital phase. Can be used to evenly sample observations or target quadratures.
        ii. Observe as often as possible subject to a minimum duration between observations.
    i. Intra-program priority. No inter-program priority ranking.

5. **Develop queue software that can be run in:**
    a. **Forecast mode.** This mode will schedule observations given a proposed distribution of nights during semester block scheduling while simulating weather losses probabilistically. The merit function would be the simulated completion fraction among all programs. This mode would account for RA pressure through its merit function. For example, if a large number of KPF users wish to observe the Kepler field at RA ~ 19 hr, the merit function will favor a distribution of nights with concentrations in Apr-Jul in A semesters.
    b. **Dispatch mode.** This mode would be run by the OA during Community Cadence blocks. During nominal operations, the OA inputs the weather band and estimates sky brightness based on current conditions. The queue returns a single target to be observed. To respond to changing weather or scheduling constraints, the OA may also filter results based on a target's *V*, RA/dec, alt/az, expected completion time.

6. **Queue software merit function should favor**
    a. A fair distribution of telescope time across programs.
    b. Low slew/acquisition overheads.
    c. A high completion rate across all programs. One issue to consider is the completion of a high cadence run. The queue must have some look-ahead capability so as to avoid scheduling small pieces of a high cadence sequence. For example, it should not schedule two isolated observations on two consecutive nights in a program that requires 10 observations within a month.

7. **Allocate KPF Community Cadence time in quarter night blocks.** The science return of KPF is maximized when the instrument is used on as many nights as feasible. We justify scheduling KPF Community Cadence in quarter-night blocks, instead of say 0.1-night increments or half-night minimums, with the following arguments. First, WMKO currently splits nights down to one quarter (although there are a few cases of smaller splits). To interface with the existing division of nights, scheduling in units of quarter night blocks is minimally disruptive. Second, quarter night scheduling will result in a Community Cadence schedule that has sufficient cadence to achieve the science goals. Given the current demand for PRV science with HIRES (a proxy for KPF demand), quarter night block scheduling will result in most nights that are not fully dark being scheduled with at least a quarter night of KPF time, as shown in the next section. Third, quarter nights are longer than the minimum duration (~2 minutes), typical duration (10-15 minutes), and maximum duration (~1 hour) of KPF exposures. Thus, observations from many KPF-CC programs can be scheduled during a given quarter night without worrying that the KPF exposures will underfill that time period. And several programs can have high-cadence observations executed concurrently at high cadence during a period of a few weeks to months. KPF Community Cadence allows for the possibility that institutions need not be required to match their own partial nights because KPF can be scheduled on a fraction of any Keck I night.

8. **Schedule a maximum of two instruments per night other than KPF.** Each instrument requires a varying level of support from the day crew and staff astronomers. Today, WMKO strongly disfavors nights split between more than two instruments. KPF will require daily calibrations, but that effort is the same whether or not the instrument is on-sky. With KPF-CC, support for observers having a variable level of experience is not needed, as is the case for other Keck instruments. The deployable tertiary also makes switching to KPF straightforward and fast. Therefore, we recommend WMKO schedule up to two instruments per night other than KPF.

9. **Full night requests for all instruments must include justifications.** To achieve the desired number of quarter night blocks allocated to Community Cadence, KPF time must be matched with ¼, ½, or ¾ night blocks from other instruments. With the current scheduling process, many Keck PIs request full nights, limiting the number of options to match KPF fractional nights. We judge that many programs can be completed with combinations of fractional nights without significantly compromising their scientific returns, although some will require full nights. This recommendation would not change the amount of time allocated for each program (e.g., a 3-night run can be split into four ¾ blocks and a 1-night run can be split into two half nights).

   We recommend that PIs of Keck I programs specify the night fractions that are *not acceptable* on their cover sheets. PIs requesting allocations composed of full nights will require a short written justification to be reviewed by the appropriate TAC or selecting official. Such justifications might be scientific, i.e. the science case can only be executed with an allocation of full nights. Other justifications might be practical. For example, the observers may not be able to observe remotely, and travel to Waimea is impractical if the allocated time is split into non-contiguous segments. With WMKO's recent support of "Pajama Mode" observing, the need to travel to Waimea is reduced even for observers without dedicated remote-obs facilities. We recognize that changing the default schedule allocation from "1 Keck night" to smaller increments is a change in the operating model of WMKO. However, we assert that it is necessary to achieve the science goals of the Observatory's latest instrument and will have minimal science impact on most non-KPF programs.

10. **Flat inter-program priority ranking.** Some queue-based implement an inter-program priority scheme where highly ranked proposals have the greatest chance of full completion. This is not possible in the Keck system where each institutional TAC reviews proposals and awards time independently. Ranking KPF programs between TACs would be problematic. We recommend continuing the current Keck policy of treating all programs with equivalent priorities.

11. **TACs to award time assuming nominal weather losses.** With a flat priority scheme, PIs should expect to share each weather band and losses of observing time equally. In other words, KPF PIs should assume their award will be *X*% of Band-1, *Y*% Band-2, *Z*% Band-3, and *W*% lost, where *X, Y, Z, W* are based on historical averages. TACs would be empowered to judge whether a proposer's plans for each band are scientifically compelling.

12. **Programs will be charged according to weather conditions at the time of observation.** For each program, the queue software will monitor the remaining time in different bands. To ensure a fair distribution of weather conditions, when an observation is taken, the weather band will be recorded into the queue database and debited from the appropriate account.

13. **Allow for target duplication.** This preserves the scientific independence of each TAC and PI. To prevent excessive duplication of effort, the metadata of each observation (target name, observation time, and exposure duration) will be published immediately to KOA. PIs would specify the proprietary period (default or custom) of downstream data products (spectra and RVs). This protocol is not required for KPF-CC, but will facilitate coordination between facilities and is consistent with the equivalent policy for NEID and the TESS follow-up working group.

14. **Keck Observatory to run and maintain Community Cadence**. Observatory staff would run the queue software and perform diagnostics to monitor the progress of programs in aggregate (e.g., completion rate and time to completion). The PI of each program is expected to be attentive to the data quality and observing cadence and provide feedback to the KPF-CC manager, if necessary

15. **Establish a Community Cadence Board of Advisors.** This board would provide guidance on queue operations and policy decisions. Participation from at least three KPF users representing different research groups would be ideal to allow for recusal in case of conflicts of interest.

16. **Formalize a system for addressing issues regarding program scheduling and completion.** We expect that most issues could be resolved by the KPF-CC manager. In exceptional cases, the WMKO Chief Scientist could make the final decisions and could solicit advice from the KPF-CC Board of Advisors.

17. **Specify how targets can be changed mid-semester.** An important aspect of Keck's existing classical observing culture is that PIs are empowered to modify their target lists up to the time of observation to be able to respond nimbly to new science opportunities or to course-correct a survey. Given that the distribution of KPF time will incorporate the proposed target coordinates, large modifications to one PI's target list may interfere with another PI's science. We recommend a middle ground where PIs may make small modifications to their target lists or observing strategies (affecting <20% of their total proposal) without review. Larger changes would require review by a WMKO official.

18. **Initiate Community Cadence as a shared-risk resource.** We acknowledge that there will be a period of learning and growth if this policy is implemented. We recommend early semesters using Community Cadence be offered as a shared-risk resource. The KPF instrument team could help support the development and transition to observatory-run Community Cadence. This could include partnering with WMKO to develop the software to support queue operations and being "on-call" to troubleshoot instrument issues related to the implementation of KPF-CC.

## Alternative Policies Considered

1. ***Ad hoc* queue run by a single research group.** This is the current mode of operations for HIRES, where the queue is run by CPS. Given the real 2021A distribution of CPS nights, Cadence Mode B has sufficiently dense sampling to model out some stellar activity signatures, although Cadence Mode C gives a higher return on investment. To be successful, PIs of small programs are required to work with a research group that manages the queue operations, e.g. California Planet Search for HIRES. The ability of a single research group to decline to work with another team disadvantages other PIs, specifically those that receive small awards of a few total nights.

2. **An observatory-managed queue operated by a pool of observers.** We believe that the OAs are the best suited to judge weather conditions and visually identify targets in crowded fields. The OAs are also not biased toward any particular program.

# Proof of Concept: Mock Schedule with Community Cadence.

In recent semesters, PRV science with HIRES has received a total of 45-48 full nights. We investigated the feasibility of scheduling a complete Keck semester according to our recommended Cadence Mode C. To review, in this mode 25% of Keck I time is distributed among 75% of all Keck I nights. To understand how such a scheme would interact with other programs, a subset of the committee (Erik Petigura, Carolyn Jordan, and Andrew Howard) created a mock 2021A semester while implementing Cadence Mode C. We explain our process below.

We divided 2021A's 181 nights into 724 quarter nights. We identified dark quarter nights where the moon was below the horizon for the entire quarter. We then made initial assignments of 181 quarter nights for KPF-CC randomly into a subset of the 724 quarter nights. The following blocks were not available to KPF-CC.

- Dark full (Q1-Q4)

- Dark first-half (Q1-Q2). However, Q3-Q4 is available.
- Dark second-half (Q3-Q4). However, Q1-Q2 is available.

We divided KPF-CC blocks as follows:

- 25% allocated in Q1 (45 nights)
- 25% allocated in Q4 (45 nights)
- 25% allocated as Q1+Q2 (23 nights)
- 25% allocated as Q3+Q4 (23 nights)

In total, 178 KPF quarter nights were scheduled on 134/181 Keck nights. The 3 quarter-night difference between 178 and 181 is a rounding effect. Figure 6 shows the distribution of KPF-CC nights over the mock 2021A semester.

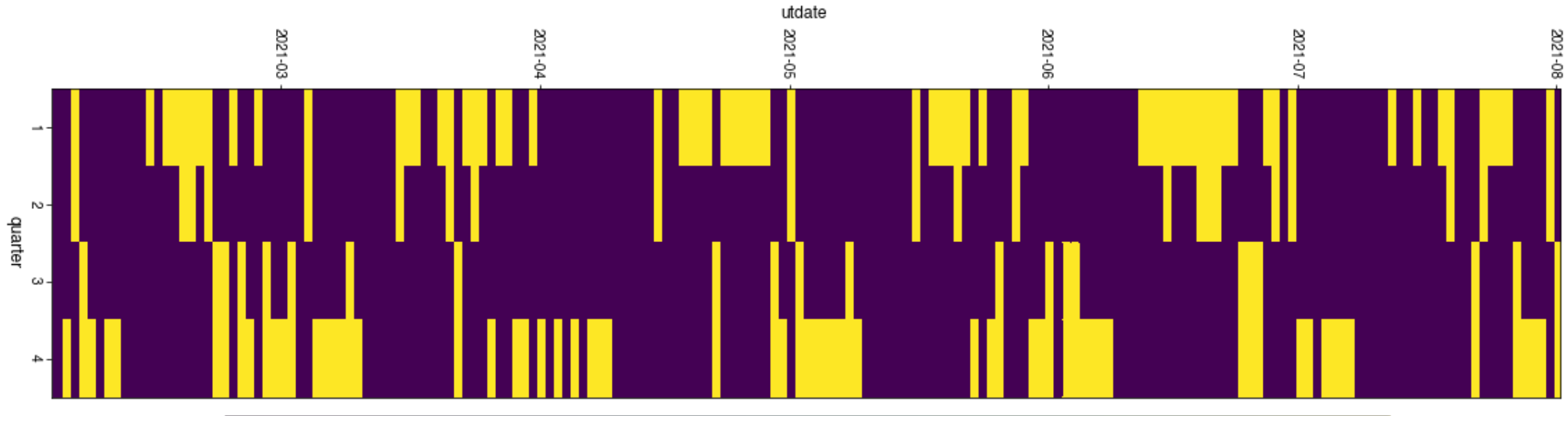

*Figure 6. Quarter nights by date over the 2021A semester. Yellow—KPF-CC assigned, Purple—not assigned.*

We uploaded the real 2021A schedule into a google spreadsheet. All nights were divided into quarters. Figure 7 shows the quarter-night assignments for the half lunation spanning 2021-05-26 to 2021-06-09. In the time period shown, there are 13 dark quarter-night blocks with HIRESr/CPS users scheduled even though dark time is not required for this science. We also uploaded the KPF-CC assignments.

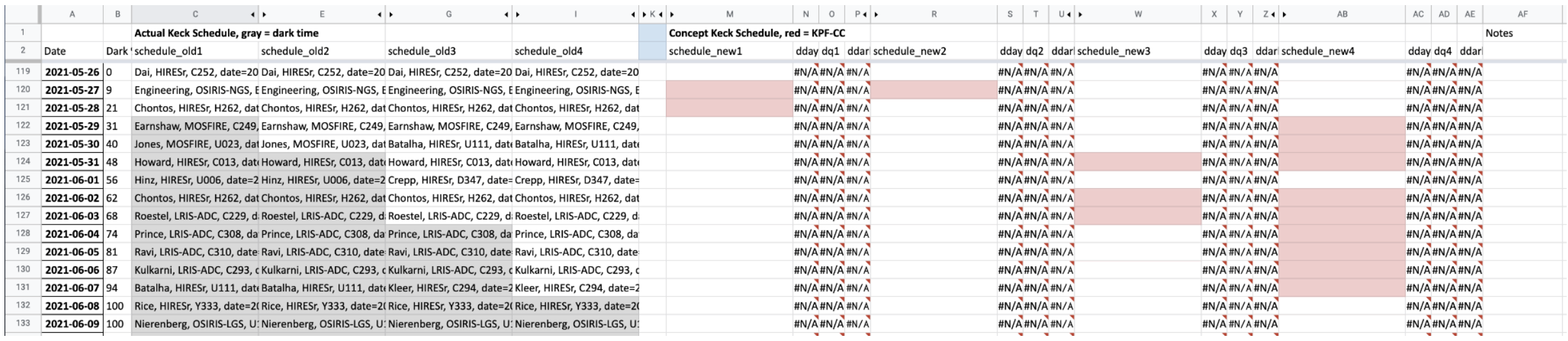

| | A | B | C | E | G | I |
|---|---|---|---|---|---|---|
| 1 | | | Actual Keck Schedule, gray = dark time | | | |
| 2 | Date | Dark ' | schedule_old1 | schedule_old2 | schedule_old3 | schedule_old4 |
| 119 | 2021-05-26 | 0 | Dai, HIRESr, C252, date=20 | Dai, HIRESr, C252, date=20 | Dai, HIRESr, C252, date=20 | Dai, HIRESr, C252, date=20 |
| 120 | 2021-05-27 | 9 | Engineering, OSIRIS-NGS, E | Engineering, OSIRIS-NGS, E | Engineering, OSIRIS-NGS, E | Engineering, OSIRIS-NGS, E |
| 121 | 2021-05-28 | 21 | Chontos, HIRESr, H262, dat | Chontos, HIRESr, H262, dat | Chontos, HIRESr, H262, dat | Chontos, HIRESr, H262, dat |
| 122 | 2021-05-29 | 31 | Earnshaw, MOSFIRE, C249, | Earnshaw, MOSFIRE, C249, | Earnshaw, MOSFIRE, C249, | Earnshaw, MOSFIRE, C249, |
| 123 | 2021-05-30 | 40 | Jones, MOSFIRE, U023, dat | Jones, MOSFIRE, U023, dat | Batalha, HIRESr, U111, date | Batalha, HIRESr, U111, date |
| 124 | 2021-05-31 | 48 | Howard, HIRESr, C013, date | Howard, HIRESr, C013, date | Howard, HIRESr, C013, date | Howard, HIRESr, C013, date |
| 125 | 2021-06-01 | 56 | Hinz, HIRESr, U006, date=2 | Hinz, HIRESr, U006, date=2 | Crepp, HIRESr, D347, date= | Crepp, HIRESr, D347, date= |
| 126 | 2021-06-02 | 62 | Chontos, HIRESr, H262, dat | Chontos, HIRESr, H262, dat | Chontos, HIRESr, H262, dat | Chontos, HIRESr, H262, dat |
| 127 | 2021-06-03 | 68 | Roestel, LRIS-ADC, C229, d | Roestel, LRIS-ADC, C229, d | Roestel, LRIS-ADC, C229, d | Roestel, LRIS-ADC, C229, d |
| 128 | 2021-06-04 | 74 | Prince, LRIS-ADC, C308, da | Prince, LRIS-ADC, C308, da | Prince, LRIS-ADC, C308, da | Prince, LRIS-ADC, C308, da |
| 129 | 2021-06-05 | 81 | Ravi, LRIS-ADC, C310, date | Ravi, LRIS-ADC, C310, date | Ravi, LRIS-ADC, C310, date | Ravi, LRIS-ADC, C310, date |
| 130 | 2021-06-06 | 87 | Kulkarni, LRIS-ADC, C293, c | Kulkarni, LRIS-ADC, C293, c | Kulkarni, LRIS-ADC, C293, c | Kulkarni, LRIS-ADC, C293, c |
| 131 | 2021-06-07 | 94 | Batalha, HIRESr, U111, date | Batalha, HIRESr, U111, date | Kleer, HIRESr, C294, date=2 | Kleer, HIRESr, C294, date=2 |
| 132 | 2021-06-08 | 100 | Rice, HIRESr, Y333, date=2( | Rice, HIRESr, Y333, date=2( | Rice, HIRESr, Y333, date=2( | Rice, HIRESr, Y333, date=2( |
| 133 | 2021-06-09 | 100 | Nierenberg, OSIRIS-LGS, U: | Nierenberg, OSIRIS-LGS, U: | Nierenberg, OSIRIS-LGS, U: | Nierenberg, OSIRIS-LGS, U: |

| | M | N | O | P | R | S | T | U | W | X | Y | Z | AB | AC | AD | AE | AF |
|---|---|---|---|---|---|---|---|---|---|---|---|---|---|---|---|---|---|
| 1 | Concept Keck Schedule, red = KPF-CC | | | | | | | | | | | | | | | | Notes |
| 2 | schedule_new1 | dday | dq1 | ddarl | schedule_new2 | dday | dq2 | ddarl | schedule_new3 | dday | dq3 | ddarl | schedule_new4 | dday | dq4 | ddarl | |
| 119 | | #N/A | #N/A | #N/A | | #N/A | #N/A | #N/A | | #N/A | #N/A | #N/A | | #N/A | #N/A | #N/A | |
| 120 | | #N/A | #N/A | #N/A | | #N/A | #N/A | #N/A | | #N/A | #N/A | #N/A | | #N/A | #N/A | #N/A | |
| 121 | | #N/A | #N/A | #N/A | | #N/A | #N/A | #N/A | | #N/A | #N/A | #N/A | | #N/A | #N/A | #N/A | |
| 122 | | #N/A | #N/A | #N/A | | #N/A | #N/A | #N/A | | #N/A | #N/A | #N/A | | #N/A | #N/A | #N/A | |
| 123 | | #N/A | #N/A | #N/A | | #N/A | #N/A | #N/A | | #N/A | #N/A | #N/A | | #N/A | #N/A | #N/A | |
| 124 | | #N/A | #N/A | #N/A | | #N/A | #N/A | #N/A | | #N/A | #N/A | #N/A | | #N/A | #N/A | #N/A | |
| 125 | | #N/A | #N/A | #N/A | | #N/A | #N/A | #N/A | | #N/A | #N/A | #N/A | | #N/A | #N/A | #N/A | |
| 126 | | #N/A | #N/A | #N/A | | #N/A | #N/A | #N/A | | #N/A | #N/A | #N/A | | #N/A | #N/A | #N/A | |
| 127 | | #N/A | #N/A | #N/A | | #N/A | #N/A | #N/A | | #N/A | #N/A | #N/A | | #N/A | #N/A | #N/A | |
| 128 | | #N/A | #N/A | #N/A | | #N/A | #N/A | #N/A | | #N/A | #N/A | #N/A | | #N/A | #N/A | #N/A | |
| 129 | | #N/A | #N/A | #N/A | | #N/A | #N/A | #N/A | | #N/A | #N/A | #N/A | | #N/A | #N/A | #N/A | |
| 130 | | #N/A | #N/A | #N/A | | #N/A | #N/A | #N/A | | #N/A | #N/A | #N/A | | #N/A | #N/A | #N/A | |
| 131 | | #N/A | #N/A | #N/A | | #N/A | #N/A | #N/A | | #N/A | #N/A | #N/A | | #N/A | #N/A | #N/A | |
| 132 | | #N/A | #N/A | #N/A | | #N/A | #N/A | #N/A | | #N/A | #N/A | #N/A | | #N/A | #N/A | #N/A | |
| 133 | | #N/A | #N/A | #N/A | | #N/A | #N/A | #N/A | | #N/A | #N/A | #N/A | | #N/A | #N/A | #N/A | |

*Figure 7. Spreadsheet used to construct mock 2021A schedule. Columns A-I show the real 2021A schedule. Awards are subdivided in by quarter nights. Quarters that are entirely dark are shaded gray. Column descriptions: A - HST date; B - dark fraction; C, E, G, I - PI, instrument, and project code assigned to Q1, Q2, Q3, Q4. Columns M-AF show the yet unscheduled mock semester. The pink cells were assigned to KPF according to the rules described above.*

Starting with the real 2021A and empty mock 2021A semester we moved quarter nights over one-by-one to fill in the entire mock 2021A semester. We programmed the spreadsheet to track the following quantities during our assignment.

- Difference in date between real and mock assignment, dday. If the new assignment is 1 day later, dday = 1.
- Difference in quarter between real and mock assignment, dq. If the new assignment is 1 quarter later, dq = 1.
- Difference in dark status, ddark. If the new assignment is dark and old is bright, ddark = 1. If the new assignment is bright and the old one is dark, ddark = -1.

When transferring over the nights, we adhered to a set of rules pertaining to:

- **Date critical assignments**. In the real schedule, there were 19 nights with date critical science or engineering. We copied these assignments over without modification.

- **Non-CPS observers.**
    - We worked to preserve dark status, i.e. ddark = 0.
    - We worked to minimize the difference in quarter and date, i.e. minimize abs(dq) and abs(dday). If large (~1 month) changes were advantageous, Carolyn Jordan judged that such a change was feasible given the program's target list.

- **CPS assignments.**
    - We copied HIRESr/CPS quarters into the pink KPF-CC cells.
    - We did not try to minimize abs(dq) and abs(dday). Large values of abs(dday) are a consequence of broader cadence coverage.
    - In the mock schedule, no KPF-CC quarters were dark, while in the real schedule there were many examples of HIRESr/CPS assignments in dark quarters. We did not work to preserve HIRESr/CPS dark time.

- **Splitting nights.** To accommodate the larger number of non-KPF ½- and ¾-night blocks in the new schedule, we split some non-CPS blocks according to
    - Splitting a full night block into two half nights on different days. (Q1 + Q2 + Q3 + Q4) → (Q1 + Q2) + (Q3 + Q4)
    - Splitting a block of two full nights into two ¾-night blocks and one ½-night block. We minimized differences in RA coverage via 2 * (Q1 + Q2 + Q3 + Q4) → 2 * (Q1 + Q2 + Q3) + 1* (Q3+Q4) or 2 * (Q2 + Q3 + Q4) + 1* (Q1+Q2).
    - Splitting a block of three full nights split into four ¾-night blocks 3 * (Q1 + Q2 + Q3 + Q4) → 2 * (Q1 + Q2 + Q3) + 2 * (Q2 + Q3 + Q4)

- **Instrument changes**
    - We scheduled no more than two instruments per night. KPF is not included in this calculation as it will require minimal nighttime support. Instrument changes are fast and automatic with the deployable tertiary on Keck I.

- In the real schedule, MOSFIRE and LRIS science are grouped together usually with MOSIRE on during bright time and LRIS on during dark time to minimize the number of instrument changes at the K1 Cassegrain focus per month. We preserved this pattern in the mock schedule.

- **KPF-CC nights.** When necessary, we moved KPF-CC nights from our original assignment. The need for this was highest during February and March, where most of the real schedule was non-CPS observers.

Figure 8 shows the same calendar dates as Figure 7 with our manual assignments applying the rules above. The full schedule is available here: https://tinyurl.com/kpf-cc-mock-schedule

| | A | B | C | E | G | I | K | M | N | O | P | R | S | T | U | W | X | Y | Z | AB | AC | AD | AE | AF |
|---|---|---|---|---|---|---|---|---|---|---|---|---|---|---|---|---|---|---|---|---|---|---|---|---|
| 1 | | | Actual Keck Schedule, gray = dark time | | | | | Concept Keck Schedule, red = KPF-CC | | | | | | | | | | | | | | | | |
| 2 | Date | Dark | schedule_old1 | schedule_old2 | schedule_old3 | schedule_old4 | | schedule_new1 | dday | dq1 | ddar | schedule_new2 | dday | dq2 | ddar | schedule_new3 | dday | dq3 | ddar | schedule_new4 | dday | dq4 | ddar | Notes |
| 119 | 2021-05-26 | 0 | Dai, HIRESr, C252, date= | Dai, HIRESr, C252, date= | Dai, HIRESr, C252, date= | Dai, HIRESr, C252, date= | | Dai, HIRESr, C252, date= | 0 | 0 | 0 | Dai, HIRESr, C252, date= | 0 | 0 | 0 | Dai, HIRESr, C252, date= | 0 | 0 | 0 | Dai, HIRESr, C252, date= | 0 | 0 | 0 | Date Critical |
| 120 | 2021-05-27 | 9 | Engineering, OSIRIS-NG | Engineering, OSIRIS-NG | Engineering, OSIRIS-NG | Engineering, OSIRIS-NG | | Batalha, HIRESr, U111, d | -11 | 0 | -1 | Batalha, HIRESr, U111, d | -21 | 0 | 0 | Engineering, OSIRIS-NG | 0 | 0 | 0 | Engineering, OSIRIS-NG | 0 | 0 | 0 | |
| 121 | 2021-05-28 | 21 | Chontos, HIRESr, H262, | Chontos, HIRESr, H262, | Chontos, HIRESr, H262, | Chontos, HIRESr, H262, | | Engineering, LRIS-ADC, I | 4 | 0 | 0 | Engineering, LRIS-ADC, I | 4 | 1 | 0 | Engineering, LRIS-ADC, I | 6 | 2 | 0 | Weiss, HIRESr, N048, da | -18 | 1 | -1 | |
| 122 | 2021-05-29 | 31 | Earnshaw, MOSFIRE, C2 | Earnshaw, MOSFIRE, C2 | Earnshaw, MOSFIRE, C2 | Earnshaw, MOSFIRE, C2 | | Earnshaw, MOSFIRE, C2 | 0 | 0 | 0 | Earnshaw, MOSFIRE, C2 | 0 | 0 | 0 | Earnshaw, MOSFIRE, C2 | 0 | 1 | 0 | Weiss, HIRESr, N048, da | -17 | 0 | -1 | |
| 123 | 2021-05-30 | 40 | Jones, MOSFIRE, U023, | Jones, MOSFIRE, U023, | Batalha, HIRESr, U111, d | Batalha, HIRESr, U111, d | | Jones, MOSFIRE, U023, | 0 | 0 | 0 | Jones, MOSFIRE, U023, | 0 | 0 | 0 | Batalha, HIRESr, U111, d | -8 | 2 | -1 | Batalha, HIRESr, U111, d | -8 | 2 | -1 | |
| 124 | 2021-05-31 | 48 | Howard, HIRESr, C013, d | Howard, HIRESr, C013, d | Howard, HIRESr, C013, d | Howard, HIRESr, C013, d | | Earnshaw, MOSFIRE, C2 | 2 | -3 | 1 | Earnshaw, MOSFIRE, C2 | 32 | -2 | 1 | Huber, HIRESr, H270, da | -10 | 2 | -1 | Huber, HIRESr, H270, da | -10 | 2 | -1 | |
| 125 | 2021-06-01 | 56 | Hinz, HIRESr, U006, date | Hinz, HIRESr, U006, date | Crepp, HIRESr, D347, da | Crepp, HIRESr, D347, da | | Shapley, MOSFIRE, N140, | 32 | -3 | 1 | Shapley, MOSFIRE, N14 | 31 | -2 | 1 | Huber, HIRESr, H270, da | -9 | 0 | -1 | Huber, HIRESr, H270, da | -9 | 0 | -1 | |
| 126 | 2021-06-02 | 62 | Chontos, HIRESr, H262, | Chontos, HIRESr, H262, | Chontos, HIRESr, H262, | Chontos, HIRESr, H262, | | Earnshaw, MOSFIRE, C2 | 34 | 0 | 0 | Earnshaw, MOSFIRE, C2 | 34 | 0 | 1 | Earnshaw, MOSFIRE, C2 | 34 | 0 | 0 | Batalha, HIRESr, U111, d | -9 | 3 | 0 | |
| 127 | 2021-06-03 | 68 | Roestel, LRIS-ADC, C229 | Roestel, LRIS-ADC, C229 | Roestel, LRIS-ADC, C229 | Roestel, LRIS-ADC, C229 | | Roestel, LRIS-ADC, C229 | 0 | 0 | 0 | Roestel, LRIS-ADC, C229 | 0 | 0 | 0 | Batalha, HIRESr, U111, d | -8 | 1 | -1 | Batalha, HIRESr, U111, d | -8 | 1 | -1 | |
| 128 | 2021-06-04 | 74 | Prince, LRIS-ADC, C308, | Prince, LRIS-ADC, C308, | Prince, LRIS-ADC, C308, | Prince, LRIS-ADC, C308, | | Prince, LRIS-ADC, C308, | 0 | 0 | 0 | Prince, LRIS-ADC, C308, | 0 | 0 | 0 | Batalha, HIRESr, U111, d | -7 | -1 | 0 | Batalha, HIRESr, U111, d | -8 | 3 | 0 | |
| 129 | 2021-06-05 | 81 | Ravi, LRIS-ADC, C310, da | Ravi, LRIS-ADC, C310, da | Ravi, LRIS-ADC, C310, da | Ravi, LRIS-ADC, C310, da | | Ravi, LRIS-ADC, C310, da | 0 | 0 | 0 | Ravi, LRIS-ADC, C310, da | 0 | 0 | 0 | Batalha, HIRESr, U111, d | -7 | 0 | 0 | Batalha, HIRESr, U111, d | -7 | 2 | -1 | |
| 130 | 2021-06-06 | 87 | Kulkarni, LRIS-ADC, C29 | Kulkarni, LRIS-ADC, C29 | Kulkarni, LRIS-ADC, C29 | Kulkarni, LRIS-ADC, C29 | | Kulkarni, LRIS-ADC, C29 | 0 | 0 | 0 | Kulkarni, LRIS-ADC, C29 | 0 | 0 | 0 | Batalha, HIRESr, U111, d | -6 | -1 | 0 | Batalha, HIRESr, U111, d | -7 | 3 | 0 | |
| 131 | 2021-06-07 | 94 | Batalha, HIRESr, U111, d | Batalha, HIRESr, U111, d | Kleer, HIRESr, C294, date | Kleer, HIRESr, C294, date | | Batalha, HIRESr, U111, d | -6 | 0 | 1 | Batalha, HIRESr, U111, d | -6 | 0 | 0 | Kleer, HIRESr, C294, date | 0 | 0 | 0 | Kleer, HIRESr, C294, date | 0 | 0 | 0 | Date Critical |
| 132 | 2021-06-08 | 100 | Rice, HIRESr, Y333, date | Rice, HIRESr, Y333, date | Rice, HIRESr, Y333, date | Rice, HIRESr, Y333, date | | Rice, HIRESr, Y333, date | 0 | 0 | 0 | Rice, HIRESr, Y333, date | 0 | 0 | 0 | Rice, HIRESr, Y333, date | 0 | 0 | 0 | Rice, HIRESr, Y333, date | 0 | 0 | 0 | Date Critical |
| 133 | 2021-06-09 | 100 | Nierenberg, OSIRIS-LGS, | Nierenberg, OSIRIS-LGS, | Nierenberg, OSIRIS-LGS, | Nierenberg, OSIRIS-LGS, | | Nierenberg, OSIRIS-LGS, | 0 | 0 | 0 | Nierenberg, OSIRIS-LGS, | 0 | 0 | 0 | Nierenberg, OSIRIS-LGS, | 0 | 0 | 0 | Nierenberg, OSIRIS-LGS, | 0 | 0 | 0 | |

*Figure 8. Same as Figure 7, but the mock schedule is filled in. Date critical nights noted, and we copied over blocks without modification. Most non-CPS quarters are scheduled during the same day and quarter (dday = 0, dq = 0). Generally, their dark time is preserved (ddark = 0) or augmented (bright Q → dark Q, ddark = 1). We did not attempt to minimize the date difference between HIRESr/CPS and KPF-CC assignments. None of the KPF-CC quarters were dark in the initial distribution of time. During our manual assignment, it was practical to assign a small number back to dark time. On the whole, there is a net transfer of HIRESr/CPS dark quarters to KPF-CC bright quarters.*

Figure 9 is a birdseye summary of the changes to non-CPS quarters. We represent shifts with lines and shorter lines correspond to smaller shifts. The median shift was 0 days, only 10% of quarters were shifted by 9 days or more. **For the non-CPS group, there was a net increase of 27 dark quarters in the mock schedule relative to the real schedule. There was also a net decrease in the number of nights with a single program from 49 to 37.**

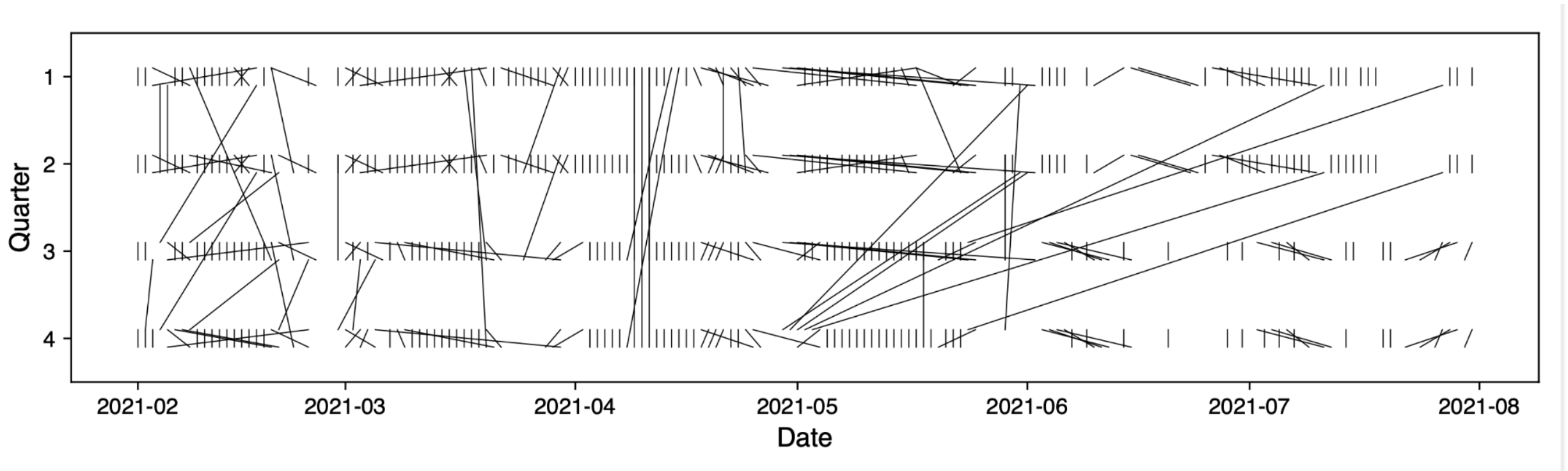

*Figure 9. Summary of changes between real and mock 2021A schedule. The night and quarter of each real non-CPS assignment are indicated by the base bottom of the lines. The top of the lines indicate the mock assignment. Most quarters are not shifted and appear as short vertical lines. Only 10% of quarters are shifted by 9+ nights.*

The following quarter nights with the largest shifts are listed below. Carolyn Jordan deemed these shifts acceptable given the program's stated scheduling constraints.

| adday | Date | schedule_new |
|---|---|---|
| 24 | 2021-05-24 | Shapley, MOSFIRE, N140, date=2021-04-30, dark=0, q=3 |
| 24 | 2021-05-25 | Shapley, MOSFIRE, N140, date=2021-05-01, dark=0, q=2 |
| 25 | 2021-03-30 | Kriek, MOSFIRE, U075, date=2021-03-05, dark=0, q=4 |
| 31 | 2021-06-01 | Shapley, MOSFIRE, N140, date=2021-05-01, dark=0, q=4 |
| 32 | 2021-05-31 | Earnshaw, MOSFIRE, C249, date=2021-04-29, dark=0, q=4 |
| 32 | 2021-06-01 | Shapley, MOSFIRE, N140, date=2021-04-30, dark=0, q=4 |
| 34 | 2021-06-02 | Earnshaw, MOSFIRE, C249, date=2021-04-29, dark=0, q=3 |
| 64 | 2021-07-27 | Ghez, OSIRIS-LGS, U017, date=2021-05-24, dark=0, q=3 |
| 69 | 2021-07-11 | Hamann, OSIRIS-LGS, U092, date=2021-05-03, dark=0, q=4 |
| 70 | 2021-07-11 | Hamann, OSIRIS-LGS, U092, date=2021-05-02, dark=0, q=4 |

Figure 10 is analogous to Figure 9 except it shows the assignment of HIRES/CPS blocks to mock KPF-CC blocks. Here, no effort was made to make small shifts since time is pooled. Compared to the HIRES/CPS in the real 2021A schedule, KPF-CC is active in 28 fewer dark quarters.

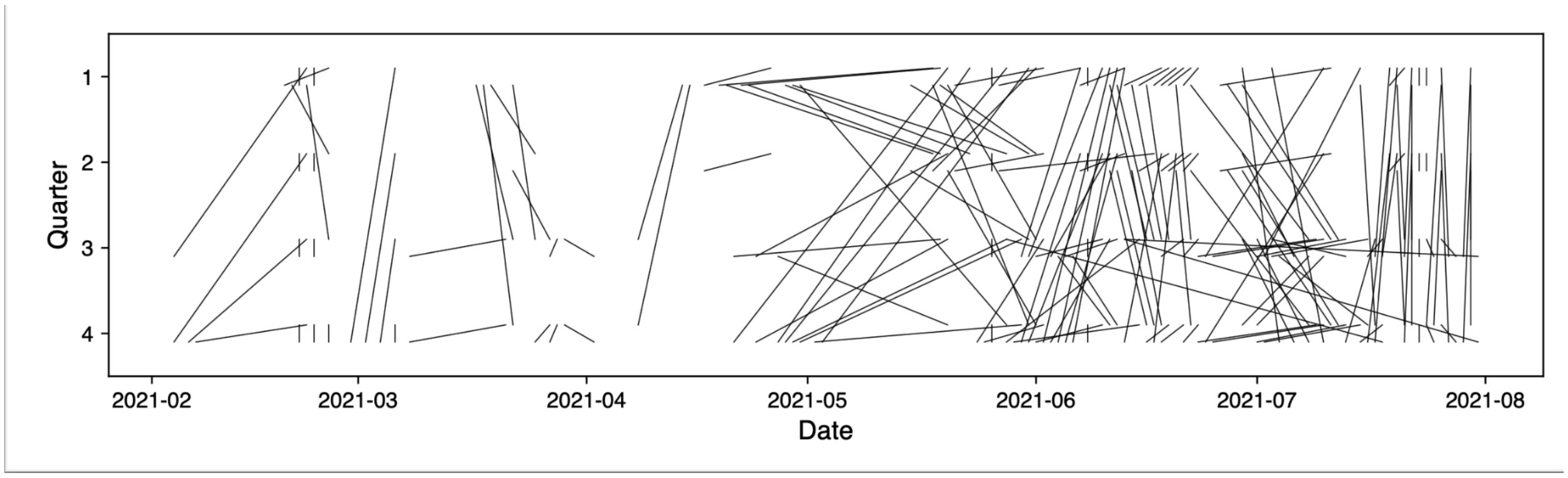

*Figure 10. Same as Figure 9 but showing the transformation of HIRESr/CPS quarters (real schedule) into KPF-CC time (mock schedule). We did not attempt to minimize date or quarter shifts and thus the changes are larger than for the non-CPS nights.*

In total, we assigned the 185 HIRES/CPS quarters nights into 100 unique nights following

- 29 Q1 or Q4 nights
- 64 Q1+Q2 or Q3+Q4 nights
- 7 Q1+Q2+Q3+Q4

The distribution of nights is somewhat different from our target distribution of 134 unique nights. The main challenge was finding 3/4 quarter nights to match with 1/4 KPF nights. Our scheme for splitting full night allocations into 3/4 nights was conservative. We only split apart consecutive blocks of 2 or more full nights. In addition, we worked to preserve the RA coverage. We believe that many PIs can make effective use of ¾ night allocations. If we were starting from scratch with a full list of ¾-night compatible proposals, we believe we could have scheduled far more ¼ night KPF blocks.

# Next Steps

This report serves to document the scientific motivations for cadence observing and the approaches taken by other observatories, to provide an existence proof for scheduling feasibility at WMKO, and to suggest specific policies to implement a robust Community Cadence program. Due to time limitations, the report does not flesh out a detailed path to implement Community Cadence. If the SSC agrees with the basic conclusions of this report, we recommend the formation of a follow-on "implementation committee". This committee could be composed of members of the committee that wrote this report, members of the KPF instrument team, and WMKO staff.

The implementation committee could be charged with constructing a detailed cost model, writing software requirements, and fleshing out details of policy recommendations. Staffing this committee may require that the Observatory allocate Keck staff hours for, e.g., DSI programmers to consider interfaces with that system and to construct conceptual designs.

## Phases of Development

We envision three phases for Community Cadence that should be planned for:

1. A **preparatory** phase (which this report has started and an implementation committee would continue). KPF will see first light in summer 2022 with expected shared-risk observations in 2022B. Completing the preparatory phase as expeditiously as possible will enable the full realization of KPF's scientific potential most quickly.
2. A **shared-risk** period of operating Community Cadence with KPF on-sky during which its implementation would be refined. This shared-risk period for Community Cadence could last a year or more as the KPF instrument comes on sky. We expect that the demands on WMKO staff time would be highest during this phase.
3. A **steady-state** operational phase when Community Cadence procedures, software, and observations would be well-established and demands on the WMKO staff would be reduced.

The software needed to implement Community Cadence is non-trivial, but examples of comparable systems exist for each component (for example, from CPS, NEID, LCO, and MINERVA).  First, a system will be needed for users to enter KPF targets and their observational parameters.  This system could also be used to track the status of proposed observations throughout the semester (as a time series) and provide links to data products that are already planned to be made available through the Keck Observatory Archive (e.g., time series of radial velocities and stellar activity indicators).  Second, a software scheduler is needed to select targets for each night of Community Cadence observing based on a set of rules and priorities.  The scheduler software can be evaluated by criteria including fairness and program completion rate using simulated data and initial observations during the shared-risk period.  Finally, Community Cadence will need to be integrated into many aspects of the Data Services Initiative (DSI), including as a graphical user interface for the KPF observing interface. It would be the job of the implementation committee to specify the requirements for these software elements.

## Costs

A key element of our cost model is that KPF observations are simple enough that in Phase 3 (steady-state operations) they can be performed by the Observing Assistants while they are operating the Keck I telescope. The implication is that additional observers do not need to be hired (or recruited from the community of KPF users) to execute the observations. This provides substantial cost savings compared to other models with professional observers. Our assertion that this model is feasible is supported by the recognition that KPF observations have very few parameters that need to be specified compared to observations with other Keck instruments. The instrument has nearly a single purpose — to measure radial velocities — and does not have general-purpose features that are typical of all other Keck instruments. KPF has only a single mode of input — one optical fiber — instead of multiple different slits/fibers, observing configurations, plate scales, etc. By design, the spectrometer has no moving parts so there are no grating angles to adjust, optics to focus, or modes to select. The parameters that specify an observation can be captured straightforwardly as a DSI observing block. These parameters include the target name/coordinates, exposure termination conditions (time-limited or SNR-limited), and the calibration strategy (simultaneous reference or bracketed reference of wavelength calibration spectrum). The calibrations for KPF spectra will be obtained automatically every afternoon and morning by a routine procedure. On-sky calibrations are not needed except in special cases and in these cases, DSI observing blocks can specify the calibrations. Thus, the job of the OA will be to push an “expose” button for a pre-selected target whose parameters have been fully specified. The idea that KPF observations can be executed by the OAs during the steady-state Phase 3, without adding a substantial burden to their jobs, is a conclusion of this committee which includes representatives from the KPF instrument team, the Lead Observing Assistant (Carolyn Jordan), and the lead Staff Astronomer involved in planning for KPF operations (Josh Walawender).  This role for the OA is similar to their role in the recent twilight observing mode.

Writing and testing the software components described above will require significant effort from professional software developers.  The first software element is an interface (and associated backend) for PIs to enter target information and track program completion throughout the semester. We estimate that this will take perhaps 1 FTE years of effort. (For this and other software elements, we offer rough estimates for the development effort but note that more reliable estimates can be determined by the implementation committee once the software is described in greater detail.) A second element is developing the scheduling code. Based on analogies with comparable projects, we estimate that developing this code will take 1-2 FTE years. The range in this estimate reflects our uncertainty because detailed requirements documents have not been written.  Finally, some of the effort to integrate DSI with Community Cadence is already part of the DSI-KPF scope of work.  Determining the scale of additional DSI effort to realize Community Cadence will require further study.

Part of the effort to develop the scheduling software will be evaluating the target lists that it generates during Phase 2. We advocate for making the scheduling software nearly automatic as the system transitions to Phase 3. This Phase 2 target list evaluation and refinement of the scheduling software may take 0.25-0.5 FTE.

The role of the Staff Astronomer (SA) on nights when KPF is scheduled will be to verify that the instrument has been calibrated properly in the afternoon (via automated scripts) and to respond to instrument and queue execution problems as they arise during the night. This is somewhat different from the SAs' role in supporting observers on other instruments who have a range of experience levels.  In some ways, SA support for KPF will be similar to their support for CPS HIRES observers, who set up that instrument independently and are largely self-sufficient during the night except when unusual instrument problems arise.  Because KPF will be scheduled on fractions of nights jointly with all of the Keck I instruments, the full set of SAs that support those instruments will need to be cross-trained on KPF observations.  This is similar to the situation for HIRES where six of the nine SAs are trained to support that instrument.

It is also worth noting changes in the operations models of KPF compared to other HIRES instruments.  For "typical" Keck instruments, the science team is responsible for taking afternoon/morning calibrations and that team "owns the night".  With KPF, calibrations will be taken automatically every night (regardless of whether the instrument is scheduled) because this calibration is needed to track and model instrument drift.  This routine calibration is needed independent of the implementation of Community Cadence.  The second change with Community Cadence is that no one KPF observer "owns the night".  While this is a change from the situation for most Keck instruments, the CPS queue operates in this mode already.  Each CPS night is formally owned by a particular PI, but this ownership does not factor into target selection.  All CPS programs are executed over the entire semester with a distribution of nights that is optimal for all of the programs.

We note that each of the costs described above is highest during the shared-risk development phase of Community Cadence, and the steady-state software maintenance will require considerably less effort.

Finally, we acknowledge that a Community Cadence program will affect the users of other Keck instruments. As described above, we believe that the scientific impacts on non-KPF users are small. We view Community Cadence in the spirit of other organizing principles of the Keck schedule that our community has come to embrace, including scheduling dark time observations where scientifically justified, date-specific observations for timed celestial events, and allowing for interrupts for target-of-opportunity observations.

## Policy Details

This report advocates for several policies that are described above. These include details about how programs will be charged for the observing time spent in a given set of conditions (seeing, transparency, etc.), whether and how metadata specifying executing KPF observations will be published, and more. The implementation committee will consider how such policies would be executed and how they will interact with the software to be developed.

# Conclusions

KPF has the potential to transform the exoplanet science from Keck Observatory. We have shown that the degree to which it meets its potential is closely linked with the observing cadence. Classically scheduled time is insufficient. *Ad hoc* queues yield lower sensitivity to planets and introduce barriers to entry and unequal access to KPF science. We believe that our proposed Community Cadence program solves these access issues. We have shown that it will enable scientific returns that are not possible with other observing cadence schemes. It is also more efficient (the result of a time series measurement can be realized more quickly). For non-KPF users, Community Cadence has the advantage of more efficiently allocating dark-time. This program does not alter the current classical mode under which most observations are scheduled. Under our proposed policy, all Keck PIs and TACs are encouraged — but not required — to accept partial nights. We provide recommendations for a path forward including an initial charge for a follow-on committee to implement Community Cadence for KPF.